\documentclass{PoS}

\usepackage{amsmath}
\usepackage{bm}
\usepackage{braket}

\newcommand{\de}{\delta}

\newcommand{\ep}{\epsilon}

\newcommand{\si}{\sigma}

\newcommand{\beq}{\begin{equation}}
\newcommand{\eeq}{\end{equation}}

\newcommand{\pa}{\partial}
\newcommand{\Tr}{\textup{Tr}}

\newcommand{\mc}[1]{\mathcal{#1}}

\title{Lorentz completion of effective string action}

\ShortTitle{Lorentz completion of effective string action}

\author{\speaker{Marco Meineri}\\
        Scuola Normale Superiore, Piazza dei Cavalieri 7, 56126 Pisa, Italy\\
        and Istituto Nazionale di Fisica Nucleare - sezione di Pisa\\
        E-mail: \email{marco.meineri@sns.it}}

\abstract{In the presence of a confining flux tube between a pair of sources the vacuum is no longer Poincar\'e invariant. This symmetry is nonlinearly realized in the effective string action. A general method for finding a large class of Lorentz invariant contributions to the action is described. The relationship between this symmetry and diffeomorphism invariance is further investigated.}

\FullConference{Xth Quark Confinement and the Hadron Spectrum,\\
		October 8-12, 2012\\
		TUM Campus Garching, Munich, Germany}

\begin{document}

\section{Introduction}
The expectation value of large Wilson loops is an order parameter for the confining phase in non-abelian gauge theories. Indeed,  a rectangular Wilson loop acts as a projector on the sector of the Hilbert space containing two non-dynamical sources, and the relative ground state energy of the sector w.r.t. the vacuum - i.e. the interquark potential - is easily extracted from the expectation value. As suggested by the phenomenon of \emph{Regge trajectories} \cite{Bali}, a leading \emph{area law} is expected to dominate the expectation value. Indeed, in this case the potential is linear in the separation, which indicates that a constant energy per unit length $\si$ is stored between the sources. The area law is in fact the leading term in a strong coupling expansion on the lattice, while higher order contributions organize themselves into a sum over surfaces with boundary on the loop. If one think of pushing the expansion all the way long to the continuum limit, one sees that the expectation value of large Wilson loops should be equal to the partition function of a two-dimensional field theory. This effective string theory describes the transverse fluctuations of a thin flux tube stretched between the sources. A first information on the action for the transverse coordinates comes again from the strong coupling expansion. Its radius of convergence does not extend until the continuum, because a \emph{roughening transition} occurs \cite{Itzykson}, which is dominated by a Gaussian fixed point. The Euclidean action in the continuum should therefore look like
\beq
\mc{S}_\textup{eff}[X]=-\si RT-\frac{b_0}{2}\int_0^T\!d\xi^0\int_0^R\!d\xi^1\,\pa_aX^i\pa_aX^i+\dots
\label{eqn:effSfree}
\eeq
Here $\xi^0,\ \xi^1$ are the space-time coordinates on the plane where the Wilson loop lies, and latin indices from the beginning of the alphabet stand for them, while the fields $X^i(\xi^a)$ describe the embedding of the worldsheet. $T$ and $R$ are the temporal and spatial extensions of the rectangular loop. The explicit integration of the partition function leads to a universal attractive correction, the so called \emph{L\"uscher term}, so that in $D$ dimensions the potential becomes \cite{Luscher:1980}
\beq
V(R)=\si R-\frac{\pi (D-2)}{24 R}+ \dots
\eeq
The constant $b_0$ contributes to the width of the flux tube, which was measured for the first time in \cite{Caselle:1996}, and it turns out that the equality $b_0=\si$ holds. This equality is readily explained if one asks for a reparametrization invariant action. The L\"uscher term has been unambiguously observed on the lattice \cite{Caselle:1997,Luscher:2002}\footnote{see also \cite{Mykkanen:2012}, where the usual multilevel algorithm \cite{Luscher:2001} is employed together with the tree-level improved action, and references therein.} and now numerical investigations can shed a light on what's behind the dots in \eqref{eqn:effSfree}. This low-energy action can be organized into a derivative expansion, and one has to rely on symmetries to fix the shape of the allowed terms. It has been recently realized \cite{Aharony:2011,Aharony:2012a} that the Poincar\'e symmetry of the underlying gauge theory provides a constructive method to tackle this problem\footnote{In the context of Galileon cosmology, the same perspective has been considered. See \cite{Nicolis,DeRham,Goon}.}. As already noticed, one should be free to choose the worldsheet parametrization, so that only a subgroup of the Poincar\'e invariant expressions should appear in the action. However, it is not clear a priori if this subgroup coincides with the local functions of the metric induced by the embedding and of the second fundamental form, because the genuine degrees of freedom are just the transverse coordinates. 

In what follows, we shall briefly discuss how the effective action for a confining string is constrained by the spontaneous Poincar\'e symmetry breaking due to the flux tube formation, and we shall then describe a systematic method for finding a large class of (in fact, possibly all) the Poincar\'e invariants \cite{Gliozzi:2012a}. In the end, the low-lying ones will be explicitly enumerated.

\section{Spontaneously broken Poincar\'e invariance}
Since the method we are meant to describe is independent of the number of dimensions and applies also to the general case of a $p$-brane, we shall keep our notation general, and we shall use Minkowski signature. The presence of a $d$-dimensional flat solitonic object in the ground state of a $D$-dimensional Poincar\'e invariant theory breaks spontaneously this symmetry down to the subgroup $ISO(1,d-1)\times SO(D-d)$. In particular, a Goldstone boson should appear for each of the broken translations, and the fields $X^i$ hold exactly this job. The symmetry allows them to appear in a low energy action. The roughening phenomenon is then understood in the continuum by means of the Mermin-Wagner-Coleman theorem \cite{Luscher:1981}. Goldstone bosons undergo a nonlinear transformation under the broken generators \cite{Coleman:1969,Callan}, but this reduces to a constant shift in the case of translations, which imposes that only field derivatives appear in the action. Since this is normally the case with Goldstone bosons anyhow, this makes little progress. However, the same fields are also the Goldstone bosons for the $d(D-d)$ broken Lorentz transformations (LT) \cite{Aharony:2011}. Indeed, it is known that the number of massless modes no longer matches the number of broken generators, in the case of broken spacetime symmetries \cite{Low}. The nonlinear realization of a LT in the plane $(b,i)$ has been found as a linearly realized LT followed by a reparametrization, in the context of a gauge fixed diffeomorphism invariant theory \cite{Goddard, Aharony:2011}:
\beq
\de^{bi}_\ep X_\mu=\ep\left(\de^b_\mu X^i-\de^i_\mu\xi^b- X^i\pa^bX_\mu\right),
\label{eqn:Nonlinall}
\eeq
where a Greek index stands for a generic coordinate. However, it is not difficult to imagine at least an action which is invariant under the transformation \eqref{eqn:Nonlinall} but not under reparametrization: a constant Lagrangian density provides this obvious example\footnote{Another example is provided by a tadpole, which could arise in the case of one transverse coordinate because $\de X$ is a total derivative. However, we look for a homogeneous ground state in which $\braket{X}=0$, so we won't consider this possibility.}. Therefore, this nonlinear transformation should be considered independently of the method used to derive it, and in the end its validity is ensured by the fact that it provides a realization of the algebra of LT. We have, for instance,
\beq
[\de^{cj}_\eta ,\de^{bi}_\ep]X_\mu=
-\ep\eta\left[\de^{ij}\left( \xi^c\de^b_\mu-\xi^b\de^c_\mu +\xi^b\pa^c X_\mu-\xi^c\pa^b X_\mu\right)+\eta^{bc}\left(X^j\de^i_\mu-X^i\de^j_\mu\right)\right].
\eeq
In \cite{Aharony:2012a} the transformation rule \eqref{eqn:Nonlinall} was systematically used for the first time to constrain the effective action. The interesting variations are the ones involving derivatives of the transverse fluctuations, which are easily seen to preserve the number of derivatives minus the number of fields. Following \cite{Aharony:2012a}, we call this the \emph{scaling}. Since terms of different \emph{scaling} are unrelated by rotational symmetry, we can organize the invariants according to their \emph{scaling}, at all orders in the derivative expansion. Moreover, since in the r.h.s. of \eqref{eqn:Nonlinall} addends with a different number of derivatives are present, the requirement of a vanishing variation of the action will lead to recurrence relations for the couplings order by order. As we are going to show, it is not difficult to sum up the associated series in a large class of cases.

\section{A diagrammatic method}
\begin{figure}[b]
\centering
\includegraphics[width=0.73\columnwidth,height=0.16\textheight]{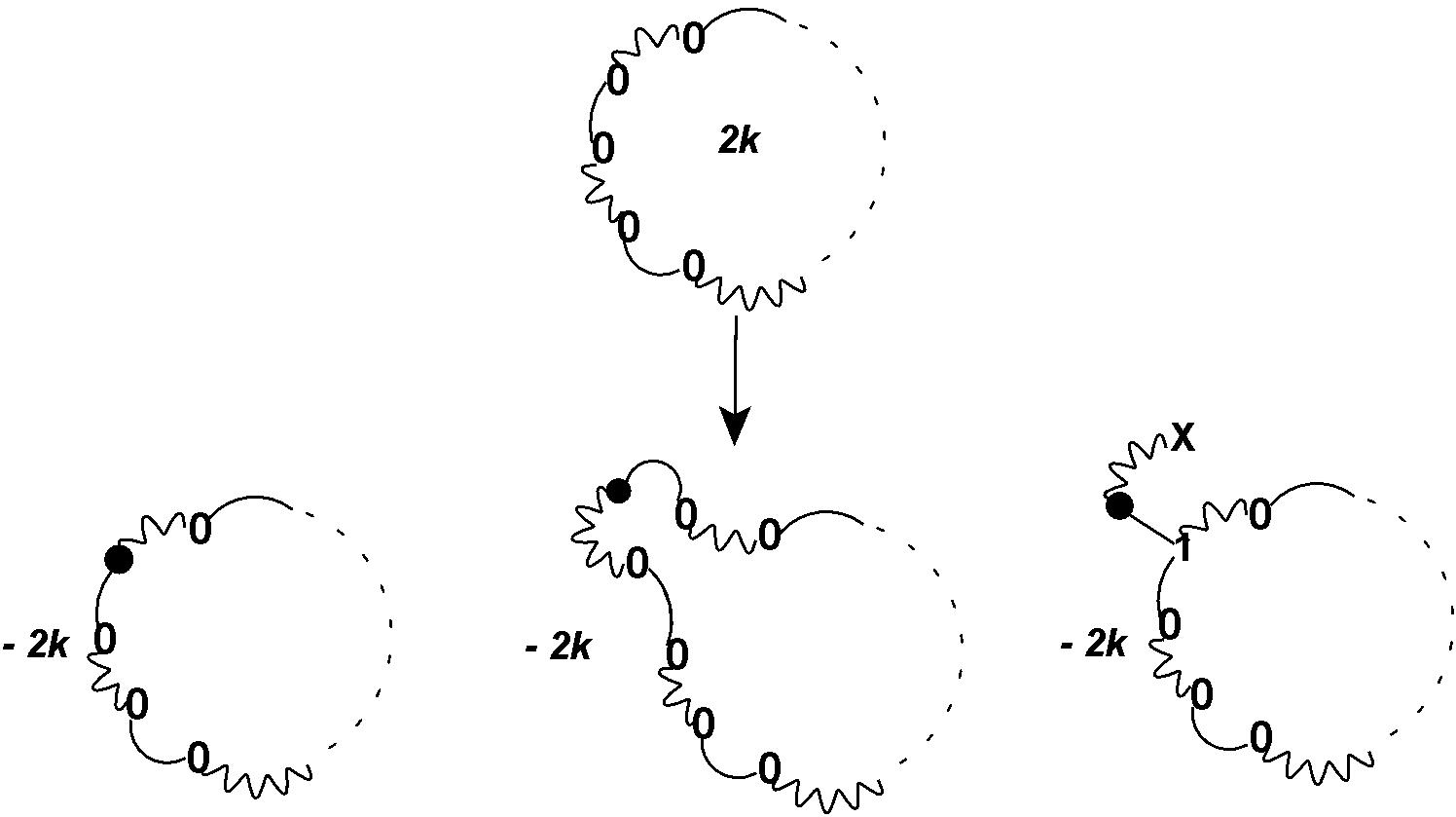}
\caption{Variation of a ring at order $2k.$ The dot represents the matrix $\ep^{aj}$ of parameters of the transformation, the $\bm{0}$s and the $\bm{1}$ are first and second derivatives respectively, while the $\bm{X}$ is the non derived field.}
\label{fig:varring}
\end{figure}

According to the symmetry breaking pattern, both worldvolume indices ($a,b,\dots$) and flavour indices ($i,j,\dots$) should be contracted. All polynomials in field derivatives obtained this way must be acted upon with the variation \eqref{eqn:Nonlinall}, and coefficients must be fixed by requiring the result to be a divergence. Let us begin by illustrating the procedure in the \emph{scaling zero} case. It is useful to write the variation of the first derivative in a covariant form, $\ep^{aj}$ being the matrix of parameters of the LT:
\beq
\de \left(\pa_bX^i\right)=-\ep^{aj}\de^{ij}\eta_{ab}-\ep^{aj}\pa_bX^j\pa_aX^i-
\ep^{aj}X^j\pa_a\pa_bX^i.
\label{eqn:cov}
\eeq
Now we associate each expression which is invariant under the stability subgroup to a graph, where the nodes represent the field derivatives and are marked by the \emph{scaling} (i.e. a node of the kind $d^mX^n$ is marked by $m-n$), and there are two kinds of link connecting the nodes. Solid lines represent saturation of worldvolume indices while wavy lines are associated to the saturation of transverse indices. At \emph{scaling zero} the only possible structure is a disconnected set of polygons with an even number of vertices, i.e. a product of traces of powers of the matrix $\pa_a X^i\pa_b X^i$. One of these \emph{rings} together with its variation is shown in fig.\ref{fig:varring}. 
It is clear by inspection that by summing \emph{rings} of increasing diameter with the appropriate coefficients the first two addends in the variation can be canceled. As for the third, it should be noticed that we get a total derivative by moving the solid link of the dot around from one vertex to the other. We therefore need to add to the Lagrangian density a copy of the first \emph{ring} times a factor $(1/2)\pa_a X^i \pa^a X^i$ to complete the total derivative. This in turns is the beginning of a new \emph{ring}, and we get another recurrence relation. It is easy to sum up the series and find the most general Lagrangian density at \emph{scaling zero}\footnote{In fact, the Cayley-Hamilton theorem introduces linear dependences among $\emph{rings}$ of different sizes in every finite dimension. This leads to linear dependencies among invariants of the same scaling, but it's not difficult to argue that no new invariant can arise because of these relations.} up to an additive constant $-\si$ \cite{Gliozzi:2012a}:
\beq
\mc{L}_0=-b_0\sqrt{-g}+b_0,\qquad
 g=\det{g_{ab}}=\det{\left(\eta_{ab}+\pa_aX^i\pa_bX^i\right)}.
\eeq
If we choose $\si=b_0$ we obtain the diffeomorphism invariant Dirac-Born-Infeld action with vanishing electromagnetic field\footnote{The case in which a photon propagates on the $p$-brane has been treated in \cite{Gliozzi:2012b}. See also \cite{Casalbuoni}.}. 

\begin{figure}[b]
\centering
\includegraphics[width=0.45\columnwidth,height=0.13\textheight]{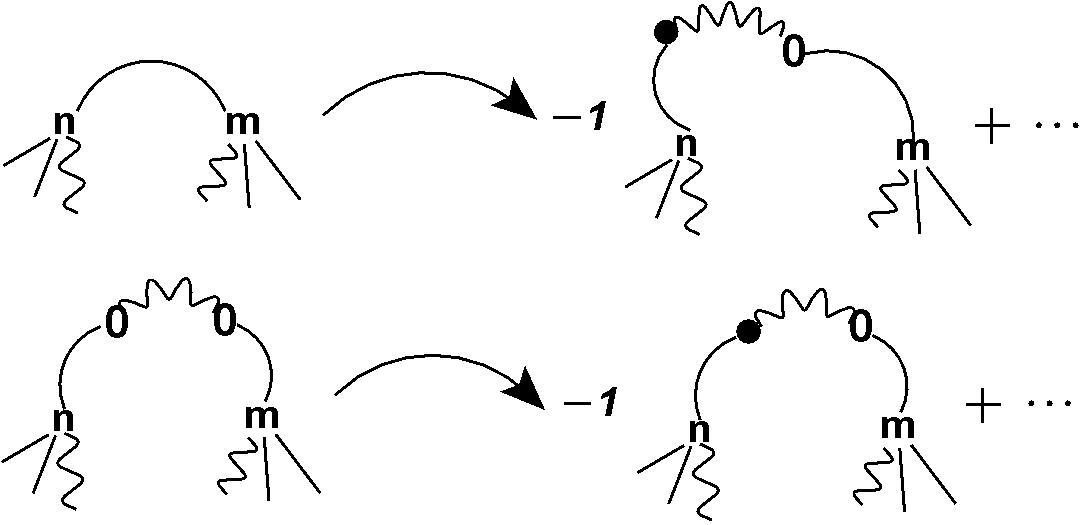}
\caption{In the first line, a variation of the vertex $\bm{n}$. In the second line, the variation of the $\bm{0}$ on the left, which equals the first one.}
\label{fig:varplain}
\end{figure}

\begin{figure}[bp]
\centering
\includegraphics[width=0.75\columnwidth,height=0.075\textheight]{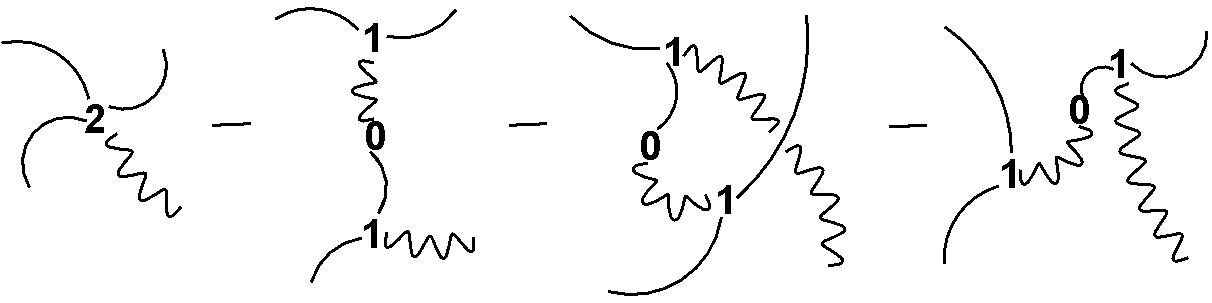}
\caption{Splitting a vertex of \emph{scaling two}.}
\label{fig:split}
\end{figure}

When moving to higher \emph{scaling}, higher derivatives enter the game. We call \emph{seeds} the graphs which do not contain \emph{scaling zero} nodes. The procedure we are going to set up permits to find all invariants whose lowest order contribution is a linear combination of \emph{seeds}. The variation of a field derivative of generic order can be expressed in a compact way by adopting the same notation explained for the graphs ($n>0$):
\beq
\de\bm{n}^i_{a_1\dots a_{n+1}}=
-\ep^{aj}\sum_{k=-1}^{n}\sum_{\{p\}}
\bm{k}^j_{a_{p(1)}\dots a_{p(k+1)}}
(\bm{n-k})^i_{a\, a_{p(k+2)}\dots a_{p(n+1)}},
\qquad \bm{-1}\equiv X,
\label{eqn:vargen}
\eeq
where the sum over $\{p\}$ counts the inequivalent ways of sharing the indices between $\bm{k}^j$ and $(\bm{n-k})^i$. First of all, consider the $k=-1$ contribution. As in the \emph{scaling zero} case, one has to add all the towers of growing \emph{rings} in order to obtain a divergence. In other words, a Lorentz invariant of scaling $n>0$ has the form $\sqrt{-g}F_n$, where $F_n$ is a suitable (still infinite) linear combination of terms of scaling $n$. Then comes the contribution of $k=0$. The diagrammatic representation of this variation, together with the structure to be added to the Lagrangian density to enforce the symmetry (fig.\ref{fig:varplain}), shows that in this case too we get a recurrence relation. By summing the associated series one obtains the following rule: \emph{in each scalar product of the worldvolume indices, replace $\eta^{ab}$ with $g^{ab}$, where}
\beq
g^{ab}=\eta^{ab}-\eta^{ac}\pa_cX^i\pa_dX^i\eta^{db}
+\eta^{ac}\pa_cX^i\pa_dX^i\eta^{de}\pa_eX^j\pa_fX^j\eta^{fb}-\dots
\eeq
\emph{is the matrix inverse of $g_{ab}$}.
The contribution of $k=n$ can be treated in a similar manner. The rule in this case is the following: \emph{in each scalar product of the transverse coordinates, replace $\delta^{ij}$ with $t^{ij}$, where} 
\beq
t^{ij}=\delta^{ij}-\pa_aX^ig^{ab}\pa_bX^j\,.
\eeq 
As for the other contributions in \eqref{eqn:vargen}, which arise only for invariants of \emph{scaling four} or higher, they correspond to all possible splittings of a node into two, whose \emph{scaling} form a partition of the original one. As usual, in the terms to be added to the Lagrangian density a \emph{scaling zero} node substitutes the dot (i.e. the matrix of coefficients of the transformation). It is easy to take into account the combinatorics in the diagrammatic representation, because the permutations in \eqref{eqn:vargen} simply correspond to all possible inequivalent ways of attaching the links to the new nodes. In fig.\ref{fig:split} an example is shown. The generalization to higher order vertices is straightforward. It is maybe worth to notice that the bulk scalar products which involve \emph{scaling zero} nodes, arising with this last procedure, should \emph{not} be modified anymore.
This completes the recipe to promote a \emph{seed}, which is invariant under the action of the stability subgroup, to an expression which is invariant under the whole Lorentz group. In the next section we shall enumerate the invariants of lowest \emph{scaling}.

\section{The Lorentz invariant action}
At \emph{scaling zero}, as already specified, the only Lorentz invariant Lagrangian densities are a constant and the volume form. In the case of the effective string, the explicit expression of the Nambu-Goto action is
\beq
\mc{L}_0=-b_0\sqrt{-g}=-b_0\sqrt{1+\Tr h+\frac{1}{2}(\Tr h)^2-\frac{1}{2}\Tr (h^2)}, \qquad 
h_{ab}=\pa_aX^i\pa_bX^i.
\eeq
At \emph{scaling two}, there are two possible \emph{seeds}, namely $(\pa_{ab}^2X\cdot \pa_{cd}^2X)\eta^{ac}\eta^{bd}$ and $(\pa_{ab}^2X\cdot \pa_{cd}^2X)\eta^{ab}\eta^{cd}$.\footnote{Notice that even if a \emph{seed} is a total derivative up to the free equations of motion, and therefore it could be perturbatively swept out by field redefinitions, we take it into account. Indeed, the Lorentz invariant Lagrangian density we obtain by applying the rules established so far needs not to be a divergence up to the free e.o.m. (but it can be a divergence up to the DBI e.o.m.). When calculating the perturbative contribution to the interquark potential, these questions should be addressed.} The associated invariants contain the two possible contractions of the squared second fundamental form ($K^i_{ab}$):\footnote{In the case of one transverse coordinate, it is easy to deduce from these expressions the existence of the invariant of \emph{scaling one} $\mc{L}_1\propto\sqrt{-g}g^{ab}K_{ab}$, which cannot be directly detected in this general case (and is not parity invariant). However, it is easy to modify the procedure in order to work directly in this context.}
\begin{align}
\mc{L}^1_2 =&c_0\sqrt{-g}\Big(\partial_{ab}^2X^k\partial_{cd}^2X^k\, g^{ac}g^{bd}
-\partial_{ab}^2X^k\partial_{cd}^2X^i\partial_eX^k
\pa_fX^i\, g^{ac}g^{bd}g^{ef}\Big)=c_0\sqrt{-g}g^{ac}g^{bd}K^i_{ab}K^i_{cd} \\
\mc{L}^2_2 =&d_0\sqrt{-g}\Big(\partial_{ab}^2X^k\partial_{cd}^2X^k g^{ab}g^{cd}
-\partial_{ab}^2X^k\partial_{cd}^2X^i\partial_eX^k
\pa_fX^i\, g^{ab}g^{cd}g^{ef}\Big)=d_0\sqrt{-g}\left(g^{ab}K^i_{ab}\right)^2.
\end{align}
Their difference is the Hilbert-Einstein action, which despite being a total derivative in the case of the effective string still plays a role in dimensional regularization \cite{Dubovsky}. It easy verified that the rules enforce the substitution $\pa_a\pa_bX^i\to\nabla_a\nabla_bX^i=K_{ab}^jn_j^i$, where $\nabla_a$ is the covariant derivative w.r.t. the induced metric and the vectors $n_j$ span the space normal to the brane in each point. Are the \emph{scaling four} corrections still all diffeomorphism invariant? They are. In fact, the output of the three rules applied to the expression $\pa_a\pa_b\pa_cX^i$ is still covariant:
\beq
\pa_a\pa_b\pa_cX^i \to \nabla_a\nabla_b\nabla_cX^i+(\nabla_b\nabla_cX^j)(\nabla_a\nabla_dX^j)g^{de}\pa_eX^i=
(\nabla_aK^j_{bc})n_j^i.
\eeq
Besides all scalars arising from traces of four copies of the second fundamental form, there is one more Lorentz invariant at \emph{scaling four}, namely
\beq
\mc{L}_4=e_0\sqrt{-g}\,\nabla_aK_{bc}^i\nabla_eK_{fg}^ig^{ae}g^{bf}g^{cg}.
\label{eqn:four}
\eeq

\section{Conclusions and perspectives}
The low energy effective action of a long confining flux tube has been studied under the only assumption of Lorentz invariance of the underlying gauge theory. From this perspective, the degrees of freedom we take into account are the Goldstone bosons of broken generators. Geometric quantities like the first and second fundamental forms are not introduced a priori, but arise as a consequence of the nonlinear realization of Lorentz algebra. When this approach was systematically introduced in \cite{Aharony:2012a}, one of the main motivations was to set up a constructive way of detecting contributions to the effective action which are \emph{not} local functions of geometric quantities, and could therefore escape a different analysis. We find no contributions of this kind up to \emph{scaling four} apart from a constant, and it would be interesting to understand if this pattern extends to all orders. We leave this to future work. A first remark is that our analysis is completely classical: Lorentz algebra is plagued by anomalies at the quantum level, therefore new issues arise, and non-invariant counter-terms have to be added to the action \cite{Dubovsky}. As for the constant, its presence in the action is irrelevant from a classical point of view. However, the interquark potential is determined by the entire partition function, therefore also this issue should be addressed from a quantum perspective \cite{Luscher:2004}. When dealing with the effective string, boundary terms have to be considered. A general method to constrain the boundary action can be set up along the same guidelines illustrated here. In fact, the first correction compatible with Lorentz invariance comes from the boundary, and has been tested on the lattice \cite{Billo,Gliozzi:2013}. It would be important also to isolate the signal of the first bulk correction to the Nambu-Goto action, which however is not allowed up to six derivatives \cite{Gliozzi:2012a,Teper:2011a,Teper:2011b}.

\end{document}